\pdfoutput=1
\documentclass{article}

\usepackage{hep-paper}

\graphics{figures}
\bibliography{bibliography}

\preprint{CP3-19-11}

\title{Heavy Neutrinos in displaced vertex searches\\at the LHC and HL-LHC}

\author{Marco Drewes \email{marco.drewes@uclouvain.be}}

\author{Jan Hajer \email{jan.hajer@uclouvain.be}}

\affiliation{Centre for Cosmology, Particle Physics and Phenomenology, Université catholique de Louvain, Louvain-la-Neuve B-1348, Belgium}

\acronym{SM}{Standard Model}
\acronym[$\nu$MSM]{nMSM}{Neutrino Minimal Standard Model}
\acronym{DM}{dark matter}
\acronym{LHC}{Large Hadron Collider}
\acronym[HL-LHC]{HLLHC}{high luminosity LHC}
\acronym{VELO}{vertex locator}
\acronym{RICH}{ring imaging Cherenkov}

\begin{document}

\maketitle

\begin{abstract}
We study the sensitivity of displaced vertex searches for heavy neutrinos produced in $W$ boson decays in the LHC detectors ATLAS, CMS and LHCb.
We also propose a new search that uses the muon chambers to detect muons from heavy neutrino decays outside the tracker.
The sensitivity estimates are based on benchmark models in which the heavy neutrinos mix exclusively with one of the three Standard Model generations.
In the most sensitive mass regime the displaced vertex searches can improve existing constraints on the mixing with the first two Standard Model generations by more than four orders of magnitude and by three orders of magnitude for the mixing with the third generation.
\end{abstract}

\section{Introduction}

Heavy right handed neutrinos $\nu_R$ appear in many extensions of the \SM of particle physics and provide an elegant explanation for the masses of the \SM neutrinos via the type-I \emph{seesaw mechanism} \cite{Minkowski:1977sc, GellMann:1980vs, Mohapatra:1979ia, Yanagida:1980xy, Schechter:1980gr, Schechter:1981cv}.
Depending on their masses, they can potentially also explain several other open puzzles in cosmology and particle physics, \cf \eg \cite{Drewes:2013gca} for an overview.
For instance, they may explain the matter-antimatter asymmetry in the early universe that is believed to be the origin of baryonic matter in the present day universe
\footnote{
A discussion of this hypothesis can \eg be found in \cite{Canetti:2012zc}.
} via \emph{leptogenesis} \cite{Fukugita:1986hr} (\cf \eg \cite{Chun:2017spz} for a recent review), compose the \DM \cite{Dodelson:1993je} (\cf \eg \cite{Adhikari:2016bei, Boyarsky:2018tvu} for recent reviews) or explain anomalies in some neutrino oscillation experiments (\cf \eg \cite{Abazajian:2012ys}).

\begin{figure}
\graphic[.5]{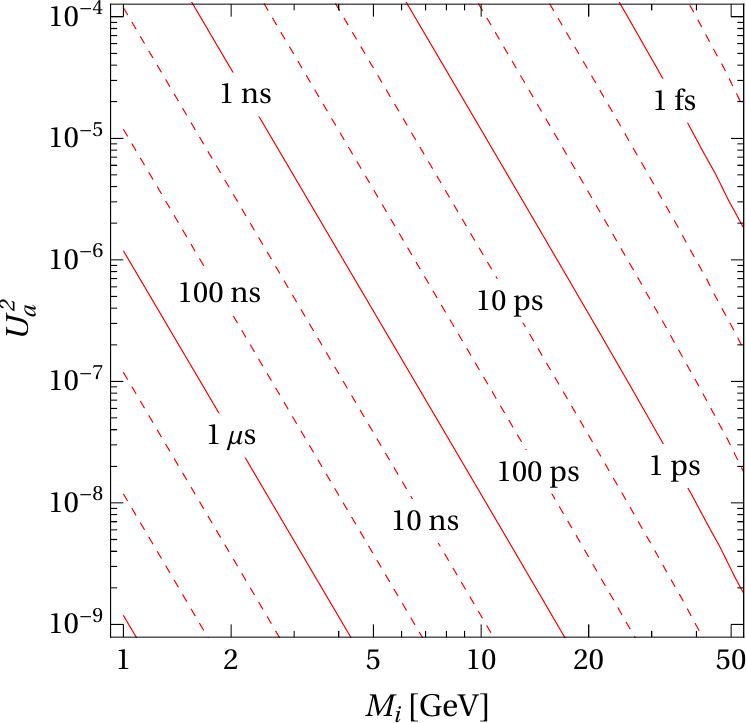}
\caption{
The heavy neutrino life time as a function of its mass and coupling calculated with \software{MadWidth} (\cf also relation \eqref{eq:Gamma}).
} \label{fig:decay length}
\end{figure}

In order to explain the light neutrino masses, the right handed neutrino flavours $\nu_{R i}$ must necessarily mix with the left handed \SM neutrinos $\nu_{L a}$, with $a = e$, $\mu$, $\tau$.
Through this mixing they can interact with the weak gauge bosons.
More precisely, the mass eigenstates $N_i$ after electroweak symmetry breaking couple to the weak interaction with amplitudes that are suppressed by small mixing angels $\theta_{a i}$ in the Lagrangian \eqref{eq:weak WW}.
This interaction is unavoidable in the framework of the seesaw mechanism.
In general the heavy neutrinos can, in addition to this, have new gauge interactions that may lead to an interesting phenomenology \cite{Cai:2017mow,Deppisch:2015qwa}.
In the present work we take the conservative approach and assume that the $N_i$ can exclusively be produced and decay via their $\theta$-suppressed weak interactions.
Then the phenomenology of heavy neutrinos $N_i$ in collider experiments can be entirely characterised by their mass $M_i$ and the mixing angle $\theta_{ai}$ that suppresses their weak interaction with \SM generation $a$:
If kinematically allowed, the $N_i$ appear in any process that involve ordinary neutrinos of flavour $a$, but with a coupling constant that is suppressed by $\theta_{ai}$ and a phase space that is affected by their mass $M_i$ \cite{Shrock:1980ct, Shrock:1981wq}.
In particular, in the relativistic regime, their production cross section is simply given by $U_{ai}^2 \sigma_{\nu_a}$, where $\sigma_{\nu_a}$ is the ordinary neutrino production cross section per generation and we have defined $U_{ai}^2 = \abs{\theta_{ai}}^2$.
Their decay width parametrically scales as $\Gamma_{N_i} \propto M_i^5 U_{i}^2$ with $U_i^2 = \sum_a U_{ai}^2$ \cite{Gorbunov:2007ak, Atre:2009rg, Canetti:2012kh, Bondarenko:2018ptm}, where the precise prefactor depends on the mass range, for the masses under consideration here it is approximately given in relation \eqref{eq:Gamma}.
\footnote{
The strong dependence of the lifetime $1/\Gamma_{N_i}$ on $M_i$ is the reason why heavy neutrinos with masses $M_i > \unit[100]{MeV}$ are not subject to any constraints from the Cosmic Microwave Background or primordial nucleosyntesis \cite{Ruchayskiy:2012si, Hernandez:2014fha}.
}
For sufficiently small mixing angles, the heavy neutrinos are \emph{long lived particles} that
can travel macroscopic distances before they decay into \SM particles, \cf \cref{fig:decay length}, giving rise to displaced vertex signatures.

\begin{figure}
\graphic[.5]{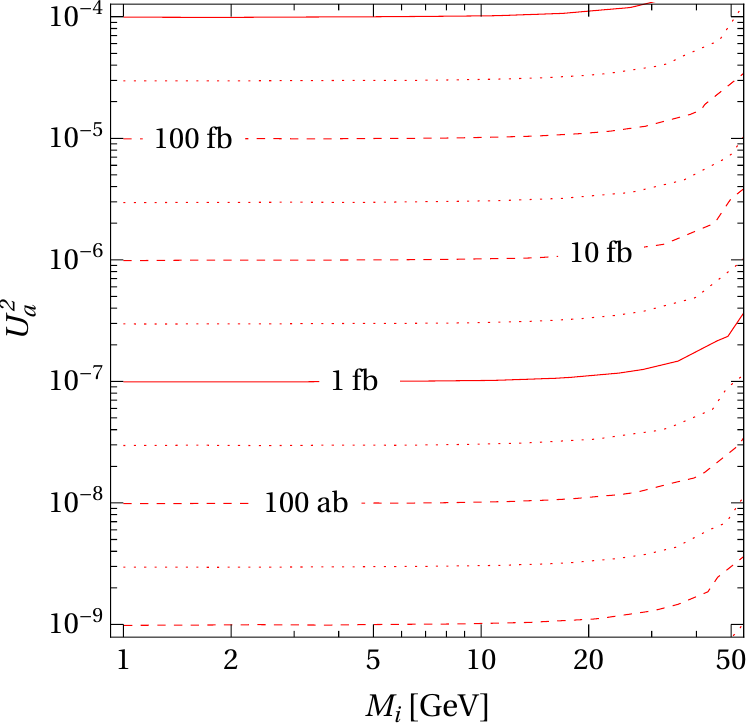}
\caption{
Cross section of the process $p p \to l_a^\pm N \to l_a^\pm l_a^\mp \overline f f^\prime$ at the \unit[14]{TeV} LHC as a function of the heavy neutrino mass and its coupling (calculated using \software{MadGraph5\_aMC@NLO}).
} \label{fig:crosssection}
\end{figure}

In the present work we explore the sensitivity of the \LHC main detectors to heavy neutrinos with masses between the $D$ meson and $W$ boson masses in the type-I seesaw model.
From a theoretical viewpoint this mass range is interesting because the heavy neutrinos could simultaneously explain the light neutrino masses via the seesaw mechanism and the matter-antimatter asymmetry of the universe via low scale leptogenesis \cite{Akhmedov:1998qx, Asaka:2005pn}, while avoiding the weak hierarchy problem \cite{Vissani:1997ys} due to the smallness of their interactions.
\footnote{
In fact, an extension of the \SM by three right handed neutrinos known as \nMSM \cite{Asaka:2005an, Asaka:2005pn} could in addition to this also explain the observed \DM density \cite{Canetti:2012vf, Canetti:2012kh}, and could in principle be a complete theory of nature up to the Planck scale \cite{Shaposhnikov:2009pv}.
}
From an experimental viewpoint this mass range is particularly interesting because the heavy neutrinos can be produced in comparably large numbers at the \LHC in the decays of real gauge bosons, \cf \cref{fig:crosssection}.
For larger masses the production cross section is much smaller than the exchanged $W$ boson is virtual, and the heavy neutrinos always decay promptly, \cf \eg reference \cite{Pascoli:2018heg} for a recent study.
For smaller masses the heavy neutrinos tend to be too long lived to decay inside the main \LHC detectors in sizeable numbers, though some improvement can be achieved by using parked data or data from heavy ion runs \cite{Drewes:2018xma, Drewes:2019vjy}.
A better sensitivity can be reached with the recently approved FASER experiment \cite{Feng:2017uoz} and other proposed dedicated detectors \cite{Chou:2016lxi, Kling:2018wct, Gligorov:2017nwh, Curtin:2018mvb, Dercks:2018wum, Alpigiani:2018fgd}, \cf \cite{Helo:2018qej, Kling:2018wct, Curtin:2018mvb}.
Fixed target experiments like NA62 \cite{CortinaGil:2017mqf, Drewes:2018gkc}, T2K \cite{Abe:2019kgx}, DUNE \cite{Krasnov:2019kdc} and in the future at SHiP \cite{Alekhin:2015byh, Anelli:2015pba, SHiP:2018xqw} are generally more sensitive than the \LHC \cite{Beacham:2019nyx}.

Currently the strongest collider constraints come from a recent CMS study \cite{Sirunyan:2018mtv, CMS:2018szz} and LEP data \cite{Abreu:1996pa}.
\footnote{
While this paper was under revision an ATLAS displaced vertex search was published that now comprises the strongest constraints in the mass region we consider here \cite{Aad:2019kiz}.
The results are consistent with our findings.
}
It has previously been pointed out in a number of theoretical papers that the sensitivity could be further improved by searching for displaced signatures \cite{Helo:2013esa, Izaguirre:2015pga, Gago:2015vma, Dib:2015oka, Dib:2016wge, Antusch:2017hhu, Cvetic:2018elt, Cottin:2018kmq, Cottin:2018nms, Abada:2018sfh, Drewes:2018xma, Drewes:2019vjy, Boiarska:2019jcw}.
We extend these studies in several ways.
\begin{enumerate}
\item We compare the sensitivity of ATLAS, CMS and LHCb in a single study.
\item We estimate the sensitivity that can be achieved with the \HLLHC in all three experiments.
Apart from an increased luminosity of \unit[3000]{\inv{fb}} for ATLAS and CMS and \unit[380]{\inv{fb}} for LHCb and collision energy of \unit[14]{TeV}, this means that CMS will be able to detect particles with larger pseudorapidity $\abs{\eta} < 4$ \cite{CMSCollaboration:2015zni}.
\item We study the sensitivity of each experiment for heavy neutrinos that mix with any of the \SM generations, using benchmark scenarios in which the mixing is exclusively with one generation at a time.
We consider both, charged and neutral current interactions, in the heavy neutrino decay.
Sensitivity estimates for heavy neutrinos that mix with $\nu_{L\tau}$ were previously made in \cite{Cottin:2018nms, Boiarska:2019jcw}, but were restricted to decays via neutral current interactions.
\item We propose a new search based on the idea to use the muon chambers for long lived particle searches \cite{Bobrovskyi:2011vx, Bobrovskyi:2012dc}, which was shown to be feasible in \cite{CMS:2015pca}.
We should add that a similar proposal was made independently in \cite{Boiarska:2019jcw}, which appeared while we were in the final stage of our analysis.
\end{enumerate}

\section{Signatures} \label{sec:signatures}

We study the \LHC sensitivity to heavy neutrinos from displaced vertex signatures in a simple model with effectively only one heavy neutrino $N$ with mass $M$ and mixing angles $\theta_a$ to the \SM flavours.
\begin{equation}
 \mathcal L
\supset
- \frac{m_W}{v} \overline N \theta^*_a \gamma^\mu e_{L a} W^+_\mu
- \frac{m_Z}{\sqrt 2 v} \overline N \theta^*_a \gamma^\mu \nu_{L a} Z_\mu
- \frac{M}{v} \theta_a h \overline{\nu_L}_\alpha N
+ \text{h.c.}
\ .\label{eq:weak WW}
\end{equation}
Here $h$ the physical Higgs field and $v \simeq \unit[174]{GeV}$ is its vacuum expectation value.
We restrict ourselves to three benchmark scenarios in each of which the heavy neutrinos exclusively mix with one \SM generation.
\footnote{
The simple model \eqref{eq:weak WW} with a single heavy neutrino is not realistic because the seesaw mechanism requires at least one flavour of right handed neutrinos to explain each non-zero light neutrino mass $m_i$, \ie, the number $n$ of their flavours $\nu_{R i}$ should be at least two ($n \geq 2$) if the lightest \SM neutrino is massless and at least three ($n \geq 3$) if it is massive.
Moreover, the heavy neutrinos must mix with all \SM generations to explain the observed light neutrino mixing angles.
Recent discussions of the constraints on the heavy neutrino flavour mixing pattern from light neutrino oscillation data can \eg be found in \cite{Drewes:2016jae, Hernandez:2015wna, Drewes:2018gkc}.
A justification for our simple phenomenological model \eqref{eq:weak WW} is given in \cref{sec:seesaw}.
}
For fixed $U^2$ we find that the sensitivity is best for $U^2 = U_\mu^2$ and worst for $U^2 = U_\tau^2$, for realistic scenarios in which the heavy neutrinos mix with several \SM generations one can expect that the number of events lies somewhere between these extremes.
The dependence of the total number of events that can realistically be observed on the flavour mixing pattern is too complicated to obtain reliable estimates by a simple interpolation.
A similar analysis in reference \cite{Drewes:2018gkc} suggests that the sensitivity is roughly given by the values that we find for the scenarios with $U^2 = U_\mu^2$ or $U^2 = U_e^2$ unless the total mixing $U^2$ is strongly dominated by $U_\tau^2$, \ie, as soon as $U_\mu^2/U^2$ or $U_e^2/U^2$ considerably exceed $\sim \unit[10]{\%}$.

\begin{figure}
\begin{panels}{.47}
\graphic{W-Decay}
\caption{
Charged current decay.
} \label{fig:charged current}
\panel{.06}
\panel{.47}
\graphic{Z-Decay}
\caption{
Neutral current decay.
} \label{fig:neutral current}
\end{panels}
\caption{
Feynman diagrams for the processes $p p \to l_a^\pm N \to l_a^\pm l_a^\mp \overline f f^\prime$ for neutrino decay via charged and neutral current in Panel \subref{fig:charged current} and \subref{fig:neutral current}, respectively.
} \label{fig:processes}
\end{figure}

We search for displaced vertex signatures from the $N$ decay inside one of the \LHC main detectors.
We consider only the dominant production through the decay of real $W$ bosons, in which the $N$ are produced along with a neutrino or charged lepton $\ell_a$, \cf \cref{fig:processes}.
We do not take the production via $Z$ decays into account because this channel does not produce a prompt charged lepton from the interaction point that one can trigger on.
The authors of reference \cite{Abada:2018sfh} have included such processes, but we estimated that the much stronger cuts that are necessary to suppress backgrounds in this case make the gain in sensitivity marginal.
For the same reason we neglect the production via Higgs decays.
At the lower end of the mass spectrum under consideration here the $N$ can also be produced in $B$ hadron decays.
We have estimated the reach of searches for displaced vertices from $N$ that are produced in $B$ decays for CMS in \cite{Drewes:2018xma, Drewes:2019vjy},
\footnote{The sensitivity has been calculated more rigorously for LHCb in \cite{Boiarska:2019jcw} and for MATHUSLA in \cite{Curtin:2018mvb}.
} here we do not include it because this would require a refinement in the computation of the production cross section from $B$ decays that goes beyond the scope of this work.
Moreover, for ATLAS and CMS the vast majority of events would fail to pass the cut on the transversal momentum $p_T$ of the prompt lepton.

For the $N$ decay we include processes mediated by both, virtual $W^*$ and $Z^*$ bosons.
For decays mediated by the charged current, a charged lepton of flavour $a$ is necessarily produced along with the $W^*$ bosons.
The $W^*$ can then decay into leptons and neutrinos or quarks that hadronise, so that the final states of the $N$ decay can be leptonic or semileptonic.
$N$ decays mediated by the neutral current can also have purely hadronic detector-visible final states.

For $a = e$, $\mu$ the first lepton produced in $W^*$ mediated decays is detector stable and can be used to reconstruct the displaced vertex.
For $a = \tau$ the $\tau$-lepton decays within the detector, mostly pions, leptons and neutrinos.
It has been pointed out in reference \cite{Cottin:2018nms, Cottin:2018nms} that the finite lifetime of the $\tau$-lepton implies that the published ATLAS efficiencies for displaced vertices \cite{Aaboud:2017iio} cannot be applied because they assume that all decay products appear promptly at the displaced vertex.
To avoid this problem the authors only included $N$ decays mediated by the neutral current.
The same strategy was also adapted in reference \cite{Boiarska:2019jcw}.
This drastically reduces the sensitivity in the scenario with $a = \tau$ when a cut on the displaced vertex invariant mass is applied because the unobservable $\nu_\tau$ that unavoidably appears in decays mediated by $Z^*$ carries away part of the energy and momentum.
In the present work we include both, $N$ decays mediated by neutral and charged currents, for all three scenarios $a = e$, $\mu$, $\tau$.
For $a = \tau$ this can be justified with two arguments.
First, for $\tau$-leptons with energies $\sim \unit[10]{GeV}$ in the laboratory frame the opening angle between decay products is so small that it is hardly noticeable that they do not promptly originate from the displaced vertex.
Second, the displaced vertex reconstruction algorithms can be improved in the future and therefore do not pose a fundamental restriction.
Such studies are \eg already under way in the CMS collaboration.

\begin{figure}
\graphic[.5]{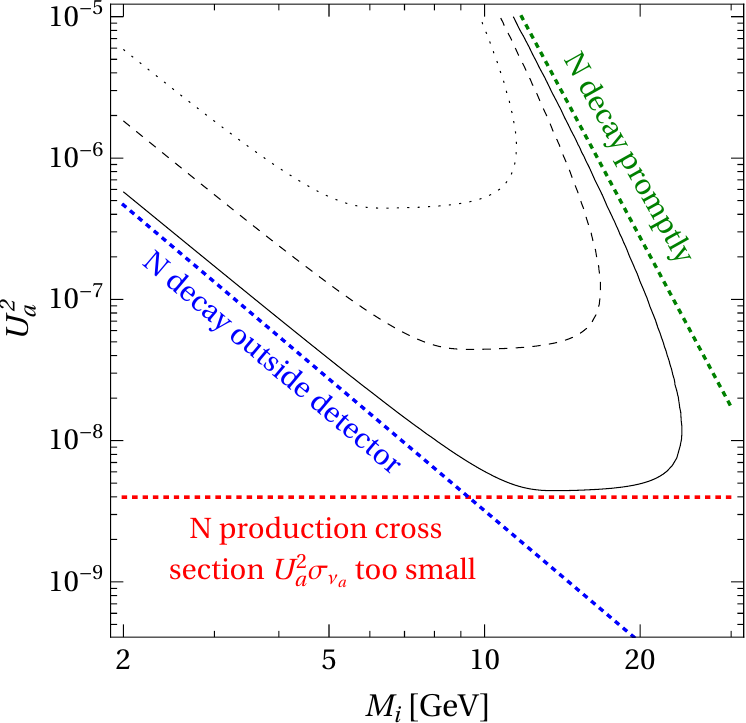}
\caption{
A simplified sensitivity estimate based on the analytic approximation \eqref{eq:number of events} using $l_0 = \unit[5]{mm}$ and $l_1 = \unit[3]{m}$ illustrates the three main obstacles in improving the sensitivity (colored dotted lines).
The three black sensitivity curves correspond to nine expected events for integrated luminosities of \unit[3, 30, 300]{\inv{fb}}, and we have assumed that all efficiencies are \unit[100]{\%}.
} \label{fig:illustrative plot}
\end{figure}

It is instructive to illustrate the dependence of the expected number of events on the model parameters with in a simplified spherical detector of radius $l_1$.
Under the assumptions summarised in \cref{sec:signatures} the cross section of events in a scenario where the right handed neutrinos mix with lepton flavour $a$ can be estimated as
\footnote{
An similar estimate for a general flavour mixing pattern is given by equation \eqref{eq:number of events2}.
}
\begin{equation}
 \sigma(W\to\ell_a N \to \ell_a \ell_b ff)
\sim
 \sigma_{\nu_a} U_a^2 \left(e^{- \gamma_N \Gamma_N l_0} - e^{- \gamma_N \Gamma_N l_1} \right)
\ ,\label{eq:number of events}
\end{equation}
where $\gamma_N$ is the Lorentz factor and $l_0$ is the minimal displacement that is required by the trigger.
An illustrative sensitivity estimate based on equation \eqref{eq:number of events} is shown in \cref{fig:illustrative plot}, it qualitatively reproduces the most important features of the results of our simulations.
For $M \gg \unit[5]{GeV}$ the $N$ decay width only depends on $U_a^2$ and $M$, it can be estimated as
\footnote{
The numerical prefactor in the decay width \eqref{eq:Gamma} depends on the way how the hadronisation of quark final states is treated.
The results given in the appendices of \cite{Gorbunov:2007ak, Canetti:2012kh} (\cf also \cite{Shuve:2016muy, Bondarenko:2018ptm}) suggest a value of 11.9, while the value used in this analysis is derived using \software{MadWidth} which neglects hadronisation effects.
}
\begin{equation}
\Gamma_N \simeq
11.9 \times \frac{G_F^2}{96 \pi^3} U_a^2 M^5
\ ,\label{eq:Gamma}
\end{equation}
see also \cref{fig:decay length}.
We recall that we only study $N$ that are produced in $W$ boson decays, \ie, $\sigma_{\nu_a}$ is the production cross section of neutrinos $\nu_{La}$ in $W$ boson decays.
We do not actually use the cross section \eqref{eq:number of events} to obtain our results, but this estimate is nevertheless helpful to qualitatively understand their dependence on the model parameters.

\section{Analysis}

We calculate the Feynman rules of heavy neutrinos coupled to the \SM with \software[2.3]{FeynRules} \cite{Alloul:2013bka} using the implementation \cite{Degrande:2016aje} based on the calculations in references \cite{Atre:2009rg, Alva:2014gxa}.
Subsequently, we generate events with \software[2.6.4]{MadGraph5\_aMC@NLO} \cite{Alwall:2011uj} (\cf \cref{fig:processes}).
Thereby, we calculate the total decay width of the heavy neutrinos with \software{MadWidth} \cite{Alwall:2014bza} and simulate their decays with \software{MadSpin} \cite{Frixione:2007zp, Artoisenet:2012st} (\cf \cref{fig:decay length}).
Finally we hadronise and shower coloured particles with \software[8.2]{Pythia} \cite{Sjostrand:2014zea}.
We calculate the efficiencies of the three \LHC main detector using our own code based on public information of the detector geometry.
\begin{itemize}
\item The ATLAS detector covers a pseudo-rapidity of $\abs{\eta} < 2.5$.
 The tracking system extends to \unit[1.1 and 3.4]{m} in the transversal and longitudinal direction, respectively, while the muon chamber covers \unit[5--10]{m} and \unit[7--21]{m} in the transversal and logitudinal direction, respectively.
\item Up to Run 3 the CMS detector covers a pseudo-rapidity of $\abs{\eta} < 2.5$, after the upgrade for the \HLLHC the pseudo-rapidity coverage will be extended to $\abs{\eta} < 4$.
The tracker extents to \unit[1.1 and 2.8]{m} in the transversal and longitudinal direction, respectively.
The muon chamber extents over \unit[4--7]{m} and \unit[7--11]{m} in the transversal and longitudinal direction, respectively.
\item The LHCb detector is optimized for measurements along the beam and has a pseudo-rapidity coverage of $2 < \eta < 5$.
It has three tracking systems, the \VELO spanning \unit[50 and 40]{cm} in transversal and longitudinal direction and two \RICH detectors.
The first \RICH is located at a distance of \unit[1]{m} from the interaction point and has a size of \unit[60 and 100]{cm} in transversal and longitudinal direction, respectively.
The second \RICH is located at \unit[9]{m} and has a size of \unit[4 and 3]{m} in transversal and longitudinal direction.
The muon chamber is located at \unit[15]{m} and has a size of \unit[5]{m} in both directions.
This layered design allows to reconstruct secondary vertices with very large displacement as the inner tracker can provide a veto for appearing track candidates.
\end{itemize}

\begin{table}
\newcommand{\pair}[2]{\ensuremath{#1}&\ensuremath{#2}}
\begin{tabular}{r*{3}{c}*{6}{r@{, }l}}
\toprule & \multicolumn{3}{c}{Single Lepton} & \multicolumn{12}{c}{Lepton Pair} \\
\cmidrule(r){2-4} \cmidrule(l){5-16} & $e$ & $\mu$ & $\tau$ & \pair{e}{e} & \pair{e}{\mu} & \pair{e}{\tau} & \pair{\mu}{\mu} & \pair{\mu}{\tau} & \pair{\tau}{\tau} \\
\midrule \multirow{2}{*}{$p_T^\text{min}$ [GeV]} & 27 & 27 & 170 & \pair{18}{18} & \pair{8}{25} & \pair{30}{18} & \pair{15}{15} & \pair{30}{15} & \pair{40}{30} \\
& & & & \multicolumn{2}{c}{} & \pair{18}{15} & \multicolumn{2}{c}{} & \pair{23}{\phantom{1}9} \\
\bottomrule
\end{tabular}
\caption{
Minimal transverse momenta used for the ATLAS single lepton and lepton pair triggers \cite{ATL-DAQ-PUB-2017-001}.
} \label{tab:triggers}
\end{table}

We propose to search in event samples which have been triggered by a single lepton or a pair of leptons \cite{ATL-DAQ-PUB-2017-001, Aaij:2017rft}.
The minimal lepton $p_T$ used for the pair triggers can be considerable softer than for the single lepton triggers, most notable when a $\tau$-lepton is involved.
For the single lepton trigger at CMS and LHCb we use $p_T(e,\:\mu,\:\tau) = \unit[30,\:25,\:140]{GeV}$ and $p_T(e,\:\mu,\:\tau) = \unit[10,\:15,\:50]{GeV}$, respectively.
Exemplarily, we show the complete trigger values used for the ATLAS detector in \cref{tab:triggers}.
Due to the lack of public lepton pair triggers for the CMS experiment we assume that the values published for the ATLAS experiment are a good estimate.
For the tracking and tagging efficiencies we use the values found in the \software[3.4.1]{DELPHES} \cite{deFavereau:2013fsa} detector cards we require all particles to have $\abs{\bm p} > \unit[5]{GeV}$ in order to allow them to escape the magnetic field, but do not apply further cuts on the particles and missing transverse energy.
The reconstruction of displaced $\tau$-leptons is not very well studied, therefore we use a simplified $\tau$-tagger based on the decay products and apply the prompt $\tau$ reconstruction efficiencies.
We assume that it is possible to search for secondary vertices stemming from decays of heavy neutrinos as long as two charged tracks with a $\Delta R > 0.1$ are detectable.
As the secondary vertex reconstruction can be performed independent of the jet reconstruction, we refrain from performing jet reconstruction and do not reduce the number of tracks through vetos.
In order to reduce the background stemming from long lived \SM hadrons we require the secondary vertices to have a minimal displacement $l_0$ of \unit[5]{mm}.
During the reconstruction of the displaced vertices we require at least two tracks with a $\Delta R > 0.1$ and an invariant mass of \unit[5]{GeV} in order to suppress further backgrounds, originating in particular from \enquote{nuclear interactions} with the detector material, which are hard to simulate \cite{Coccaro:2016lnz, Alimena:2019zri}.
In the case of purely muonic displaced vertices we do not apply this invariant mass cut because the backgrounds are much lower than in the case with hadrons in the final state.
In this case it is favourable to carefully simulate the backgrounds instead of loosing signal events due to a hard cut.
Such detailed simulation goes beyond the scope of this work, and our strategy is to provide an estimate of what sensitivity could in principle be achieved in this way.
The displaced vertex reconstruction is well established if the produced particles traverse almost the entire tracker.
If a particle transverses only a part of the tracking system the efficiency must drop off.
In order to be able to compare different experiments we require that the remaining tracks transverse at least half of the tracker and that the reconstruction efficiency drops off linearly before that.
This functional form is consistent with efficiency dependence published by ATLAS \cite{Aaboud:2017iio}, and we assume that the behaviour is similar for CMS and LHCb.
In order to calculate the remaining path of the particle we use ray tracing \cite{Smits:1998ei, Williams:2005ae}.

These assumptions are slightly more optimistic than what is done in present searches.
We \eg use less tracks to reconstruct the displaced vertex than reference \cite{Aaboud:2017iio} (\cf also \cite{Cottin:2018kmq}), and we assume that the largest observable displacement can be improved by a factor of two compared to current ATLAS estimates \cite{Aaboud:2017iio}.
Our assumption that the displaced vertex invariant mass cut can be removed in the purely muonic search is consistent with reference \cite{Aad:2015uaa}, where, however, the pairing of displaced vertices could be used to reduce the background.
It can further be justified by arguing that interactions with the detector material are unlikely to create a purely muonic final state.
The main problem is that the simulation of some backgrounds, including the \enquote{nuclear interactions} with the detector material, is very difficult and requires detailed knowledge of the detector, which goes beyond the scope of this work.
Our current goal is to estimate what can be achieved if one is only limited by the properties of the detectors, \eg by using improved algorithms in the future, under reasonable assumptions.

\begin{figure}
\graphic[.5]{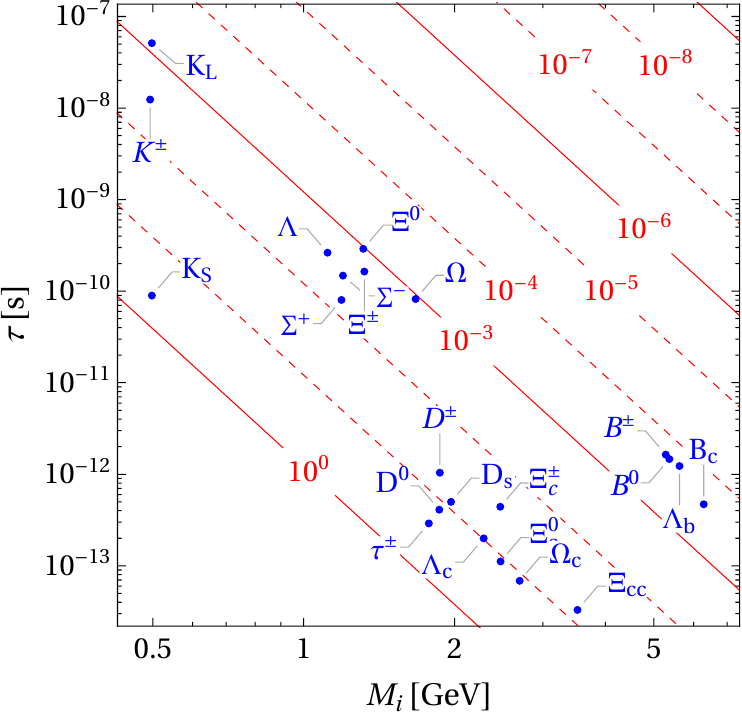}
\caption{
Heavy neutrino coupling (red lines) compared to potentially relevant \SM backgrounds (blue dots) as function of mass and life time, \cf also \cref{fig:decay length}.
} \label{fig:background}
\end{figure}

The muon chamber systems can record muons which are an order of magnitude further away from the primary interaction vertex than the inner tracking system.
Therefore, it is compelling to search for long lived particles using also the muon chambers to identify displaced vertices as we have also proposed before in the context of displaced signatures in supersymmetric models \cite{Bobrovskyi:2011vx, Bobrovskyi:2012dc}.
In the meantime it has been demonstrated that such a search is feasible \cite{CMS:2015pca}.
In order to be conservative we require reconstructed particle to transverses the whole muon chamber.

For $n$ observed events the significance is given by \cite{Cowan:2010js}
\begin{align}
 S(n | h)
&= \sqrt{- 2 \ln \frac{P(n | h)}{P(n | n)}}
 & \text{where} &
 & P(n | h)
&= \frac{h^n}{n!} \inv[h]{e}
\ ,
\end{align}
is the Poisson probability and $h$ is either the number of events predicted by the \emph{background only hypothesis} $b$ or by the \emph{background with signal hypothesis} $b + s$.
For the exclusion and discovery of a model we require $S(n | b + s) \geq 2$ and $S(n | b) \geq 5$, respectively.
In order to project to future searches we estimate the observed number of events $n$ with the prediction for the alternative hypothesis.
Hence, we use $S(b | b + s) \geq 2$ and $S(b + s | b) \geq 5$, for exclusion and discovery, respectively.
In doing so we neglect the systematic uncertainties of the signal and the background estimation.

The \SM background can be efficiently excluded with the cuts on the invariant mass and the displacement, \cf \cref{fig:background}.
It is not easy to quantify the remaining backgrounds without an extremely realistic simulation of the whole detector.
In this analysis we ignore backgrounds originating from cosmic rays and beam-halo muons, based on the low rate in the \LHC experimental caverns and the good capability of the experiments to recognize them as such \cite{Liu:2007ad}.
Furthermore, we do not include pile up in our analysis.
We also ignore the scattering of ordinary neutrinos that come from the collision point, based on the low cross section of charged-current interaction in the detector material.
We assume that the background number is smaller than one and calculate the efficiencies based on one background event.
In this case the non-observation of four events and the observation of nine events suffices for exclusion and discovery, respectively.

\section{Results}

\begin{figure}
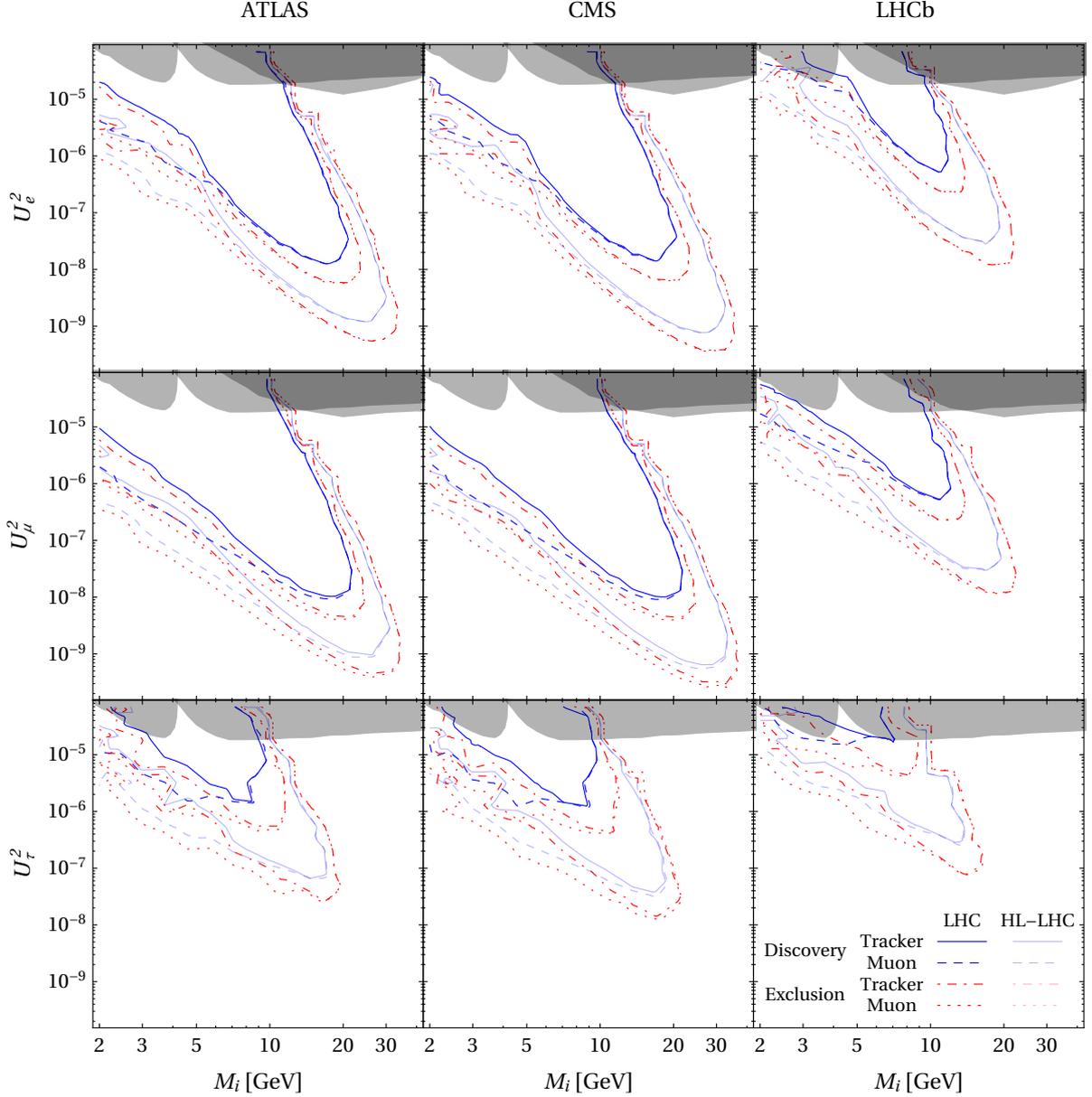

\graphic{comparison}
\caption{
Sensitivity reach of the \unit[13 and 14]{TeV} \LHC.
The three columns show the results for ATLAS, CMS and LHCb.
The luminosities assumed for CMS and ATLAS are \unit[0.3 and 3]{\inv{ab}} and for LHCb are \unit[30 and 380]{\inv{fb}}.
The three rows constitute the three extreme cases of pure electron, muon and tau coupling, respectively.
The blue (red) curves show the discovery (exclusion) potential, both of which are shown for searches using only the tracker and using the tracker together with the muon chamber.
For the simulation of LHCb we require decays to happen before or within the first \RICH (\cf also \cref{fig:LHCb}).
The gray bands represent the exclusion bound from DELPHI \cite{Abreu:1996pa} and CMS \cite{Sirunyan:2018mtv}.
} \label{fig:results}
\end{figure}

We present our results for each experiment (ATLAS, CMS and LHCb) and each benchmark scenario (exclusive mixing with $e$, $\mu$ or $\tau$) in \cref{fig:results}.
Due to their similar geometry ATLAS and CMS have comparable sensitivities, the different $\eta$ coverage in the \HLLHC Runs does not have a strong impact on the sensitivity curves because of the strong dependence of the number of events on $U^2$.
The geometry of the muon chambers is different for both detectors, but the effect of this on our results is small because we only require the heavy neutrino to decay before the muon chamber, so that the difference is practically captured by the $\eta$ coverage.
For pure electron or muon mixing they can exclude heavy neutrinos with couplings as small as $\sim 5 \times \inv[8]{10}$ and masses up to $\sim \unit[20]{GeV}$ with an integrated luminosity of \unit[300]{\inv{fb}}.
With \unit[3000]{\inv{fb}} mixing angles of $\sim 5 \times 10^{-10}$ and masses up to $\sim \unit[40]{GeV}$ become accessible.
LHCb can exclude down to $5 \times 10^{-7}$ and $5 \times 10^{-8}$ and masses up to \unit[12 and 20]{GeV} for \LHC and \HLLHC.
ATLAS and CMS can exclude heavy neutrinos with pure $\tau$-couplings up to $\sim 10^{-6}$ and $\sim 10^{-8}$ with masses up to \unit[15 and 25]{GeV} for integrated luminosities of \unit[300]{\inv{fb}} and \unit[3000]{\inv{fb}}, respectively.
The lower sensitivity of LHCb is a result of both, the lower integrated luminosity and the different $\eta$ coverage of LHCb, where the dominant effect comes from the luminosities.
For the mixing with $\tau$ this is partly compensated by the lower $p_T$ thresholds in the triggers.

\begin{figure}
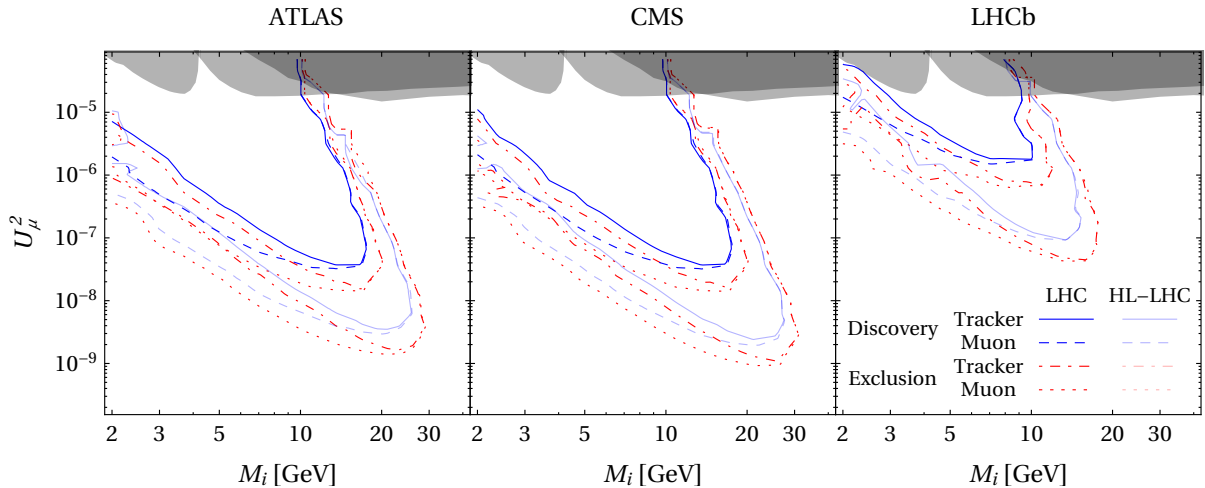

\graphic{leptonic}
\caption{
Sensitivity reach of a purely leptonic search.
} \label{fig:leptonic}
\end{figure}

\begin{figure}
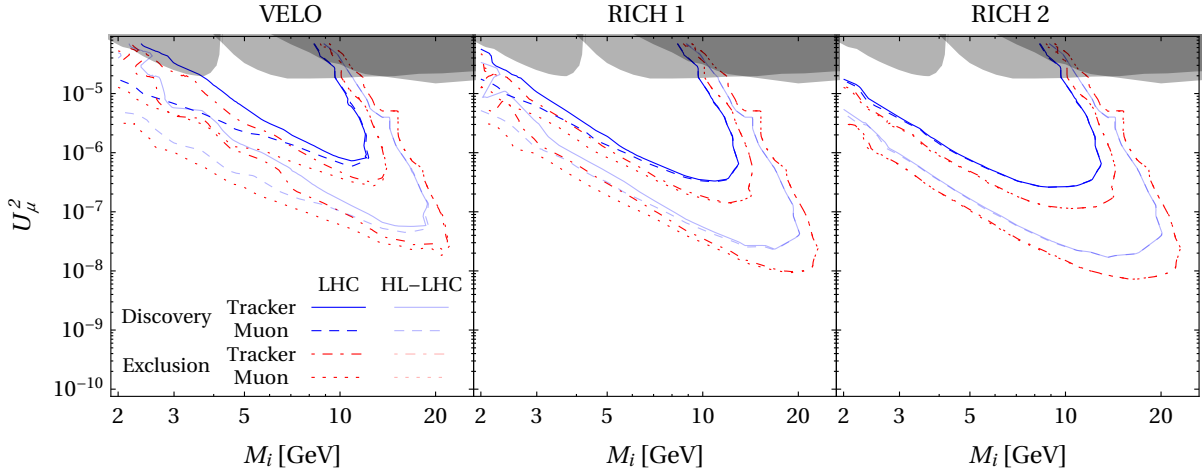

\graphic{LHCb}
\caption{
Comparison of the sensitivity reach of LHCb at \unit[13 and 14]{TeV} with \unit[30 and 380]{\inv{fb}} for pure muon mixing, under the assumptions that the neutrino has to decay within the \VELO and before or within the first or second \RICH, respectively.
} \label{fig:LHCb}
\end{figure}

For masses below the mass of optimal sensitivity the sensitivity is limited by the fact that the $N$ are too long lived and do not decay inside the detector, \cf \cref{fig:illustrative plot}.
For the mixing with electrons the displaced vertex invariant mass cut is clearly visible in \cref{fig:results}, it explains the drop in sensitivity around \unit[5]{GeV}.
The sensitivity for $M < \unit[5]{GeV}$ comes entirely from muons in the final state, for which we have removed the displaced vertex invariant mass cut.
\footnote{
The main reason why a displaced vertex invariant mass cut has been imposed in previous analyses is that there are backgrounds in this regime that were not fully quantified, including \SM backgrounds from long lived resonances like $J/\psi$ and interactions with the detector material.
However, in the currently ongoing CMS search for heavy neutrinos, no displaced vertex invariant mass cut is imposed because the \SM backgrounds can be measured directly from data down to very low dilepton masses \cite{Martina}.}
In the scenarios with pure muon or tau mixing the sensitivity that can be achieved from those is good enough that the effect of the cut is barely visible in the plots.

For long lifetimes the reach can be extended by using the muon chamber.
This approach is especially fruitful for masses as small as a few GeV because we assume that for purely muonic vertices no invariant mass cut is necessary in order to remove the hadronic background.
For $M < \unit[5]{GeV}$ the production of heavy neutrinos from meson decays dominates over the gauge boson decays considered here, hence we can expect that our proposal to use the muon chambers yields a bigger improvement in searches for those.
In the present analysis the improvement for ATLAS and CMS amounts only to a factor of order one in the scenario with pure muon mixing and less than that in the other scenarios.
This can be understood as a result of the scaling $\sigma(W \to \ell_a N \to \ell_a \ell_b ff) \propto U_a^4 M^5$ in this regime, which can be obtained by expanding the exponential functions in relation \eqref{eq:number of events} to linear order in $\Gamma_N l_0$ and $\Gamma_N l_1$ and using the estimate \eqref{eq:Gamma}.

\begin{figure}
\begin{panels}{2}
\graphic{Ue}
\caption{
Pure electron mixing.
} \label{fig:Ue}
\panel
\graphic{Umu}
\caption{
Pure muon mixing.
} \label{fig:Umu}
\panel
\graphic{Utau}
\caption{
Pure tau mixing.
} \label{fig:Utau}
\panel
\end{panels}
\caption{
Exclusion reach of ATLAS, CMS with \unit[3]{\inv{ab}} and LHCb with \unit[380]{\inv{fb}} for the \HLLHC under ideal conditions
for pure electron, muon and tau mixing in Panel \subref{fig:Ue}, \subref{fig:Umu} and \subref{fig:Utau}, respectively.
} \label{fig:summary}
\end{figure}

Additionally, we show the sensitivity reach of a purely leptonic search in the case of pure muon mixing in \cref{fig:leptonic}.
The reach in $U^2_\mu$ is one order of magnitude weaker than the inclusive reach shown in \cref{fig:results}.
Finally, we show the reach of the LHCb detector under the assumptions that the secondary vertex must either lie within the \VELO or is allowed to lie before or within one of the two \RICHs, in \cref{fig:LHCb}.

In \cref{fig:summary} we summarise our most optimistic estimates for the reach of each experiment in each of the benchmark scenarios.
For this plot we have made the same assumptions as for the most optimistic scenario presented in \cref{fig:results}, and additionally relaxed the lower bound on the displaced vertex invariant mass to \unit[2]{GeV}.
For LHCb we have in addition assumed that the \RICH 2 can be used.
This relaxation may be justified by a better understanding of the detector backgrounds in the future.
Qualitatively all of the presented results are in good agreement with the analytic estimate \eqref{eq:number of events} plotted in \cref{fig:illustrative plot}, which can explain the overall shape of the sensitivity regions.

\section{Discussion and conclusions}

We have compared the sensitivity of the \LHC detectors ATLAS, CMS and LHCb to heavy neutrinos from $W$ boson decays in displaced vertex searches in three benchmark models, in each of which the heavy neutrinos mix exclusively with one \SM generation.
Moreover, we propose a new search strategy that includes the muon chambers to detect tracks from the displaced vertex.
We summarise our main results in \cref{fig:summary}.
In \cref{fig:results} we present more detailed estimates for each of the different scenarios we investigated when both, leptonic hand hadronic final states are included.
\cref{fig:leptonic} shows the sensitivity for purely leptonic decays, and \cref{fig:LHCb} illustrates how the LHCb sensitivity depends on the part of the tracker that is used for the displaced vertex search.

We find that the sensitivity that can be reached at ATLAS or CMS with \unit[300]{\inv{fb}} exceeds existing bounds by three orders of magnitude for mixing with the first and second generation and by one order of magnitude for mixing with the third generation.
The reach of LHCb shown in \cref{fig:results} is more than an order of magnitude worse than ATLAS and CMS.
The main reason for this is the lower integrated luminosity.
However, \cref{fig:summary} suggests that the LHCb sensitivity in the scenario with exclusive coupling to the third generation can be competitive with ATLAS and CMS if the \RICH 2 is used.
Our results in this regard are not conclusive considering that we have made simplifying assumptions about the efficiencies in the \RICH 2.
Additionally, there are statistical uncertainties in this particular channel due to the comparably small number of total events.
We propose to address this issue with a dedicated study.

It is noteworthy that the sensitivity in terms of the squared mixing angle $U_a^2$ scales with the square root of the integrated luminosity for small values of $U_a^2$, which can be illustrated by expanding the simple estimate \eqref{eq:number of events} in $\Gamma_N l_1$.
The reason is that the \LHC is practically used as an intensity frontier machine.
This has the important implication that the \HLLHC can further improve the sensitivity considerably, as shown in \cref{fig:results}.
Our results qualitatively agree with other recent studies, but are more general in the sense that we for the first time compare all three detectors in all three scenarios for both, the \LHC and \HLLHC.

The considerable improvement in sensitivity compared to the existing bounds is particularly interesting from a cosmological viewpoint because it implies that the \LHC can probe a considerable part of the parameter region where low scale leptogenesis can generate the observed matter-antimatter asymmetry of the universe.
In the minimal model with two heavy neutrinos ($n = 2$) and in the \nMSM this parameter region is inaccessible to conventional \LHC searches, \cf \cite{Hernandez:2016kel, Antusch:2017pkq, Eijima:2018qke, Boiarska:2019jcw} for updated parameters scans.
In this model, independent measurements of all $U_{ai}^2$ and the heavy neutrino mass spectrum at colliders would, together with a measurement of leptonic CP violation in neutrino oscillation experiments, in principle allow to constrain all parameters in the Lagrangian \eqref{eq:Lagrangian}, making this a fully testable model of neutrino masses and baryogenesis \cite{Drewes:2016jae}.
In practice it is unlikely that data from the \LHC will be accurate enough to pin down all parameters, but measurements of the flavour mixing pattern at the \LHC would nevertheless provide an important test of the hypothesis that heavy neutrinos are responsible for the light neutrino masses and baryogenesis and motivate further studies at future colliders \cite{Antusch:2017pkq}.
If three heavy neutrinos contribute to leptogenesis and the seesaw mechanism ($n = 3$) then the viable parameter space for leptogenesis is much larger \cite{Drewes:2012ma, Canetti:2014dka}, in the mass range considered here it entirely covers the experimentally allowed region in the mass-mixing plane (white area in our plots) \cite{Abada:2018oly}.
In our simulation we find that the \LHC could possibly observe tens of thousands of events for the largest experimentally allowed mixing angles.

Our results provide a strong motivation to improve the displaced vertex reconstruction efficiencies in \LHC experiments.
Moreover, a better understanding of the backgrounds would also be highly desirable.
If one could, for instance, lower the cut on the displaced vertex mass, this would make it possible to search for heavy neutrinos with smaller masses.
Those could be produced in considerable numbers in the decay of $B$ hadrons, which are much more numerous in \LHC collisions than the $W$ bosons considered here.
In this regime the sensitivity can further be improved by using the muon chambers to detect tracks from displaced vertices that lie outside the main tracker.

\subsection*{Acknowledgments}

We are very grateful to Martino Borsato, Giacomo Bruno, Claudio Caputo, Giovanna Cottin, Didar Dobur, Andrea Giammanco, Elena Graverini, Shih-Chieh Hsu, Philippe Mermod, Maksym Ovchynnikov, Jessica Prisciandaro, Albert de Roeck and Michael Winn for comments on numerous experimental aspects.
This work was partly supported by the Fonds de la Recherche Scientifique de Belgique (F.R.S.-FNRS) under the Excellence of Science (EoS) project \no{30820817} (be.h).
Computational resources have been provided by the F.R.S.-FNRS Consortium des Équipements de Calcul Intensif (CÉCI), funded by the under Grant \no{2.5020.11} and by the Walloon Region.

\appendix

\section{Type I seesaw model} \label{sec:seesaw}

The type-I seesaw model is given by the most general renormalisable extension of the \SM Lagrangian $\mathcal L_{\SM}$ that only contains $n$ sterile neutrinos $\nu_{R i}$ as new fields,
\begin{equation}
 \mathcal L
 = \mathcal L_{\SM} + \i \overline{\nu_R}_i \slashed \partial \nu_{Ri}
 - \frac{1}{2} \left( \overline{\nu_R^c}_i M_i \nu_R + \overline{\nu_R}_i M_i \nu_{R i}^c \right)
 - F_{ai} \overline \ell_a \varepsilon \phi^* \nu_{Ri}
 - F_{ai}^* \overline{\nu_R}_i \phi^T \varepsilon^\dagger \ell_a
\ .\label{eq:Lagrangian}
\end{equation}
Here $\ell_a$ are the \SM lepton doublets and $\phi$ the \SM Higgs doublet.
The $F_{ai}$ are Yukawa coupling constants, and we have chosen a flavour basis where the Majorana mass matrix $M_M$ of the $\nu_{R i}$ is diagonal.
After the spontaneous breaking of the electroweak symmetry by the Higgs vacuum expectation value $v \simeq \unit[174]{GeV}$ there are $3 + n$ mass eigenstates:
three light neutrinos $\text{nu}_i$ with masses $m_i^2$ given by the eigenvalues of the matrix $m_\nu^\dagger m_\nu$ with $m_\nu = - \theta M_M \theta^T$ and $n$ heavy mass eigenstates $N_i$ with masses $M_i^2$ approximately given by the eigenvalues of $M_M^\dagger M_M$,
\begin{align}
\text{\nu} &= U_\nu^\dagger \left(\nu_L - \theta \nu_R^c\right) + \text{c.c.}
\ , &
N &\simeq U_N^\dagger \left( \nu_R + \theta^T \nu_L^c\right) + \text{c.c.}
\ .
\end{align}
Here c.c.\ refers to the charge conjugation, which acts as $\nu_R^c = C \overline{\nu_R}^T$ with $C = \i \gamma_2 \gamma_0$.
$U_\nu$ is the standard light neutrino mixing matrix and $U_N$ its equivalent amongst the heavy neutrinos.
We use the tree level relation and expand to leading order in the mixing between left and right handed neutrinos, which is quantified by the matrix $\theta = v F \inv{M_M}$.

Naively one would expect the magnitude of the mixing angles $\theta_{ai}$ to be comparable to the ratio between light and heavy neutrino masses, $\theta^2\sim m_i/M_i$.
However, if the $M_i$ and $F_{ai}$ approximately respect a generalised $B - L$ symmetry \cite{Shaposhnikov:2006nn}, then the mixings $U_{ai}^2 = \abs{\theta_{ai}}^2$ can be large enough to be within reach of the \LHC \cite{Kersten:2007vk}.
\footnote{
For $n = 3$ the approximate $B - L$ symmetry implies $F_{a2} \simeq - \i F_{a3}$, $\abs{F_{a1}} \ll \abs{F_{a2}}\simeq \:\abs{F_{a3}}$ and $M_2 \simeq M_3$, \cf \eg \cite{Abada:2018oly} for a detailed discussion.
In the \nMSM this could explain the feeble coupling of the \DM candidate as well as the mass degeneracy $\abs{M_3 - M_2} \ll M_2$, $M_3$ that is required to generate the matter-antimatter asymmetry \cite{Shaposhnikov:2006nn}.
Similar arguments can be made for $n > 3$ \cite{Moffat:2017feq}.
}
While the symmetry naturally explains the smallness of the light neutrino masses for below the electroweak scale and Yukawa couplings that are larger than that of the electron, it also parametrically suppresses the rate of lepton number violating processes \cite{Kersten:2007vk}.
This includes as the decay of $W$ bosons into same sign dileptons, which are considered to be a \enquote{golden channel} for heavy neutrino searches, \cf \eg \cite{Sirunyan:2018xiv, Aaboud:2018spl} for recent updates.
A precise quantification of this suppression difficult because it depends on the splitting between the heavy neutrino masses.
If all mass splittings are bigger than the experimental resolution, then cross section \eqref{eq:number of events} can simply be generalised to the case with several heavy neutrino flavours; the cross section of events with lepton flavour $a$ from the first vertex ($N$ production) and $b$ from the second vertex ($N$ decay) that can be seen in a detector can be estimated as
\begin{equation}
 \sigma(W\to\ell_a N_i \to \ell_a \ell_b ff)
 \sim
 \sigma_{\nu_a} U_{ai}^2 \frac{U_{bi}^2}{U_i^2} \left(e^{- \gamma_{N_i} \Gamma_{N_i} l_0} - e^{- \gamma_{N_i} \Gamma_{N_i} l_1} \right)
\ .\label{eq:number of events2}
\end{equation}
Parametrically this estimate applies to both, decays that violate the \SM lepton number $L$ and those that conserve it.
If the mass splittings are too small to be experimentally resolved, but still much larger than the $\Gamma_{N_i}$, then one has to simply add the event rates from the decays of the mass-degenerate states for both, $L$ conserving and $L$ violating decays.
In the context of realistic seesaw models, we can identify $U_a^2 = \abs{\theta_a}^2$ in the estimate \eqref{eq:number of events} based on the phenomenological model \eqref{eq:weak WW} with $U_{a2}^2$ or $U_{a3}^2$ in \eqref{eq:number of events}.
If the mass splitting is smaller than $\Gamma_{N_i}$, then the destructive interference between contributions from different $N_i$ to the diagram in \cref{fig:processes} leads to a suppression of processes that violate $L$ \cite{Kersten:2007vk}.
In the intermediate regime where $\abs{M_i - M_j} \sim \Gamma_{N_i}, \Gamma_{N_j}$ there can be non-trivial effects due to coherent oscillations among the heavy neutrinos \cite{Anamiati:2016uxp, Dib:2016wge,Das:2017hmg, Antusch:2017ebe, Antusch:2017pkq, Cvetic:2018elt, Hernandez:2018cgc}.
Hence, a precise prediction of the number of events depends on parameters that may not be directly observable if the mass splitting is smaller than the kinematic resolution of the detector.
However, our present analysis does not require lepton number violation because we use the displacement of the vertex in with the $N_i$ decay to suppress the \SM backgrounds.
Whether or not lepton number violating processes contribute therefore at most amounts to a factor two in the heavy neutrino decay rate $\Gamma_{N_i}$.
This rate in principle enters the cross section \eqref{eq:number of events2} exponentially.
However, in most of the testable region in the mass-mixing plane the exponentials can be approximated linearly.
In the double-logarithmic sensitivity plots in the mass-mixing plane the main difference is a shift of the sensitivity region in the direction of comparable large masses and large couplings, where it is cut off due to the fact that the $N_i$ decay before they travel a distance $l_0$.
Moreover, in the range of masses and couplings considered here, the suppression of lepton number violating processes only occurs for fine-tuned parameter choices \cite{Drewes:2019byd}.
Therefore we can perform the analysis in the effective model \eqref{eq:weak WW}.

\printbibliography

\end{document}